\documentclass{article}

\usepackage{amsmath}
\usepackage{natbib}
\usepackage{booktabs} % For pretty tables
\usepackage[left=1in, right=1in, top=1in, bottom=1in, lines=25]{geometry}
\usepackage{setspace}
\def\bSig\mathbf{\Sigma}

\usepackage[figuresright]{rotating}

\title{Generalised mixed effects models for changepoint analysis of biomedical time series data}

\author
{Mark B. Fiecas\\%\emailx{mfiecas@umn.edu}, \\
Division of Biostatistics and Health Data Sciences,\\ University of Minnesota,\\ Minneapolis, Minnesota, U.S.A.
\and
Kathryn R. Cullen\\
Department of Psychiatry and Behavioral Sciences,\\ University of Minnesota,\\ Minneapolis, Minnesota, U.S.A.
\and
Rebecca Killick \\
School of Mathematical Sciences,\\ Lancaster University,\\ Lancaster, United Kingdom, LA1 4YF
}

\begin{document}

%  This will produce the submission and review information that appears
%  right after the reference section.  Of course, it will be unknown when
%  you submit your paper, so you can either leave this out or put in 
%  sample dates (these will have no effect on the fate of your paper in the
%  review process!)

\date{}

%  These options will count the number of pages and provide volume
%  and date information in the upper left hand corner of the top of the 
%  first page as in published papers.  The \pagerange command will only
%  work if you place the command \label{firstpage} near the beginning
%  of the document and \label{lastpage} at the end of the document, as we
%  have done in this template.

%  Again, putting a volume number and date is for your own amusement and
%  has no bearing on what actually happens to your paper!  

%\volume{64}
%\pubyear{2008}
%\artmonth{December}

%  The \doi command is where the DOI for your paper would be placed should it
%  be published.  Again, if you make one up and stick it here, it means 
%  nothing!

% \doi{}

%  This label and the label ``lastpage'' are used by the \pagerange
%  command above to give the page range for the article.  You may have 
%  to process the document twice to get this to match up with what you 
%  expect.  When using the referee option, this will not count the pages
%  with tables and figures.  

\label{firstpage}

\maketitle

%  put the summary for your paper here
\begin{abstract}
Motivated by two distinct types of biomedical time series data, digital health monitoring and neuroimaging, we develop a novel approach for changepoint analysis that uses a generalised linear mixed model framework. The generalised linear mixed model framework lets us incorporate structure that is usually present in biomedical time series data. We embed the mixed model in a dynamic programming algorithm for detecting multiple changepoints in the fMRI data. We evaluate the performance of our proposed method across several scenarios using simulations. Finally, we show the utility of our proposed method on our two distinct motivating applications.
\end{abstract}

%  Please place your key words in alphabetical order, separated
%  by semicolons, with the first letter of the first word capitalized,
%  and a period at the end of the list.
%

%\begin{keywords}
%segmentation; covariance matrix; mean shift; mixed model
%\end{keywords}

%  As usual, the \maketitle command creates the title and author/affiliations
%  display 

%  If you are using the referee option, a new page, numbered page 1, will
%  start after the summary and keywords.  The page numbers thus count the
%  number of pages of your manuscript in the preferred submission style.
%  Remember, ``Normally, regular papers exceeding 25 pages and Reader Reaction 
%  papers exceeding 12 pages in (the preferred style) will be returned to 
%  the authors without review. The page limit includes acknowledgements, 
%  references, and appendices, but not tables and figures. The page count does 
%  not include the title page and abstract. A maximum of six (6) tables or 
%  figures combined is often required.''

%  You may now place the substance of your manuscript here.  Please use
%  the \section, \subsection, etc commands as described in the user guide.
%  Please use \label and \ref commands to cross-reference sections, equations,
%  tables, figures, etc.
%
%  Please DO NOT attempt to reformat the style of equation numbering!
%  For that matter, please do not attempt to redefine anything!

\section{Introduction}
% Opening 
Technological innovations in the biomedical sciences have led to an increase in the amount of and the quality of time series data collected from individuals. Two distinct examples of biomedical time series data we use in this paper are 1) data from at-home devices in digital health and 2) brain scans from a neuroimaging study on non-suicidal self-injury (NSSI). In the former, using time series data from a longitudinal study, we are interested in using activity monitoring data to model people's routine across two months, and identify subtle departures from their routine. In the latter, we are interested in using resting-state functional magnetic resonance imaging (rs-fMRI) data, which is represented by a multivariate time series for each individual, to quantify the strength of interactions between different regions of the brain, and investigate if the changes in these interactions can be used as a biomarker for NSSI. In this paper, we are interested in identifying the times for which the properties of the time series data change, also known as changepoint analysis. The identification of these shifts and changes in the properties of the time series data provide insights into human behavior or the temporal dynamics of brain activity. Our main contribution in this article is the development of generalised linear mixed effects models for changepoint analysis of time series data. Though the two examples are very distinct, the same modelling framework can be adapted in a manner that leverages the structure present in the biomedical time series data.

% Changepoint literature review
Changepoint analysis has a long history and has been applied to time series data from many scientific fields, from climate time series to financial time series \citep{Truong20}. Generally, these methods were developed to identify changes in the mean, variance, or regression coefficients for continuous-valued time series \citep{PELT,Fryzlewicz14,Shi22,zeileis02}. There have also been some developments for Poisson and binomial data \citep{Franke12,Pignatiello01}. These methodological developments allowed practitioners to use changepoint analysis across many scientific disciplines, from studying the climate \citep{Lund23}, to patterns in human home activity \citep{Taylor21}, to brain connectivity \citep{cribben13}, and to closing prices of stock market indices \citep{Cho15}, among many others. However, methodological developments are fairly limited for biomedical time series data. Biomedical time series data generally contains structure, and this can affect the properties of the data. For instance, in our first example on data from at-home devices, the study had a longitudinal design, and so the time series data has two sources of correlation: the autocorrelation over time within a series and the within-individual correlation from the repeated measures. In our second example on data using rs-fMRI time series, the data is multivariate, and the different dimensions of the rs-fMRI data form an interconnected community structure \citep{yang24}. Accounting for the structure in the time series data is critical to accurately characterise the phenomenon of interest, and will improve performance in estimation and statistical inference.
% Biomedical time series data, however, can be non-Gaussian, multivariate, and can contain structure. 
% Talk about changepoints for longitudinal regression models with random effects, but note that the changepoints are random, so the problem is not the same. Emphasize the idea of their use of random effects 
% Random effects in general are good for population-level / group level etc, which the longitudinal folks have looked into. We use a similar strategy for our purposes for time series data.
% GLMERs have been around, and we can use existing machinery to do changepoint analysis. However, we need to do a whole lot more development for covariance matrices.
% This is a general framework for both first and second order, but a lot of what's already there is first order

% Changepoints for longitudinal data
%The mixed model framework has been successfully used for changepoint analysis for analysing longitudinal data. 
The mixed model framework is one approach to account for the structure present in the data. Random effects can be carefully specified to model the structure in the data. For instance, in our first example, random effects can account for the repeated measures within an individual. In our second example, the covariance structure of the random effects can model the interconnected community structure. Prior works have successfully used random effects in changepoint models to account for the longitudinal structure in the data \citep{Dominicus08,Lai14,Naumova01}. These developments, however, were made only for longitudinal models for the mean, and thus, are not applicable to time series data. These developments nevertheless provides a foundation for using random effects for changepoint analysis, and so we will extend these methods to time series data. Random effects can also be used to parameterize the interconnected community structure in multivariate biomedical time series data \citep{yang24}. We will build on and extend this approach to changepoint analysis. Altogether, the mixed model framework is a flexible modeling framework for changepoint analysis of biomedical time series data. We will show how the careful specification of the random effects will result in a model that can account for the structure in the data.

% Summary
The rest of this article is organized as follows. Section \ref{sec:methods} describes the mixed model framework we will use for modelling changepoints, with details for the generalised linear mixed effects regression (GLMER) model, linear mixed effects covariance (LMEC) model, and the Pruned Exact Linear Time (PELT) algorithm for changepoint estimation. In Section \ref{sec:sim} we use simulations to show the performance of the GLMER and LMEC models. In Section \ref{sec:app}, we show two distinct applications of our models: at-home activity monitoring and resting-state functional magnetic resonance imaging (rs-fMRI). Section \ref{sec:discussion} concludes the article with a discussion.

\section{Methods}\label{sec:methods}
%(Common notation)
%$y_{t,i,j}$, where $t$ is time, $i$ is group, $j$ is $j$-th series in group $i$, and $\epsilon$ is the independent error.
%LMER: $y_{t,i,j} = \mu_t + \mu_{t,i} + \epsilon_{t,i,j}$; $\epsilon_{t,i,j}$ has moments that do not change over time;
%GLMER: logit$(E(y_{t,i,j)}) = \mu_t + \mu_{t,i}$; $\epsilon_{t,i,j}$ has moments that do not change over time;
%LMEC: $\mathbf{y}_{t,i} = \mathbf{\mu}_{t,i} + \mathbf{\epsilon}_{t_i}$, where $\mathbf{\mu}_{t} \sim MVN(\mathbf{0},\Sigma_\mu)$ and $\mathbf{\epsilon}_{t,i} \sim MVN(\mathbf{0},\Sigma_\epsilon)$. (Form of Sigmas here.)

In the general setting, we consider a collection of time series $\{y_{t,i,j}\}$ observed at times $t=1,\ldots,n$, for each group $i=1,\ldots,K$ where the $i$-th group contains $k_i$ time series indexed as $j=1,\ldots,k_i$.  We assume there are $M$ ordered changepoints at times $0=\tau_0, \tau_1, \ldots, \tau_M, \tau_{M+1}=n$ that partition the data into $M+1$ segments.  Each segment is assumed to follow a common distributional form but with parameters, particularly the means or the covariance matrices, that differ across segments. In the following subsections, we develop mixed effect models for changepoint analysis of a univariate or multivariate time series.

\subsection{Generalised Linear Mixed Effects Regression Model}\label{sec:glme}
In many applications, the mean behaviour can be split into a global mean, $\mu$ and a group specific mean, $\mu_i$.  The simplest form of a linear mixed effects model is,
\begin{align}
E(y_{i,j}) = \mu + \mu_{i} + \epsilon_{i,j},
\label{eqn:lme}
\end{align}
where $\epsilon_{i,j}$ is iid across $i, j$ with mean zero and variance $\sigma_{i}^2$ and the random effects $\{\mu_i\}$ are mutually independent of the error terms $\{\epsilon_{i,j}\}$.  Note that we have dropped the subscript $t$ as we consider the model within a specific segment.

Generalising this across exponential family distributions and incorporating external regressors we can write the Generalised Linear Mixed Effects Regression (GLMER) model as,
\begin{align}
    y_{i}|b \sim \textrm{Distr}\left(\mu_{i},\sigma^2\right) \\
    g(\mu)=X\beta + Zb, \label{eqn:glme}
\end{align}
where Distr is the conditional distribution, $X$ are the external regressors with associated estimated coefficients $\beta$, and $Z$ are the random effects with estimated coefficients $b$.  To recover equation \eqref{eqn:lme} we set $X$ to be a vector of 1's, $Z$ to be a matrix where each column is a 0-1 variable indicating the grouping, and $g(\cdot)$ to be the identity link.

In our simulations and applications, we consider the cases where Distr is Normal (with identity link function) or Bernoulli (with logit link function).  As we fit these to each segment, we keep the model form ($X$ and $Z$) the same and only the estimates $\beta$ and $b$ can vary across segments.

\subsection{Linear Mixed Effects Covariance Model}\label{sec:lmec}
We now consider the situation where we are interested primarily in the covariance matrix of the data. First, we arrange our data as follows. Let $\mathbf{y} = (y_{1,1}, y_{1,2},\ldots,y_{1,k_1},y_{2,1},\ldots y_{K,k_K})$ be a $n\times \sum_ik_i$ matrix of our time series data, recalling that each $y_{i,j}$ is a time series of length $n$. Let $\mathbf{\mu} = (\mu_1,\ldots,\mu_K)^\prime$ with $K\times K$ covariance matrix $\Sigma_\mu = \text{Cov}(\mathbf{\mu}) = (\sigma_{\mu,uv})$ and $\mathbf{\epsilon} = (\epsilon_{11},\ldots,\epsilon_{Kk_K})^\prime$ with diagonal $\sum_i k_i \times \sum_i k_i$ covariance matrix $\Sigma_\epsilon = \text{Cov}(\mathbf{\epsilon}) = \text{diag}(\sigma^2_{\epsilon,11},\sigma^2_{\epsilon,12},\ldots,\sigma^2_{\epsilon,1k_1},\sigma^2_{\epsilon,21},\ldots,\sigma^2_{\epsilon,Kk_K})$. The covariance matrix of the data, $\Sigma$, is thus a function of $\Sigma_\mu$ and $\Sigma_\epsilon$. The assumptions we make about the within-group and between-group covariance structure leads to different structures for the covariance matrix of the data. We focus on two structures: i) the Uniform-Block structure, and ii) the Heterogeneous-Block structure.

The Uniform-Block (UB) structure was developed by \citet{yang24}, motivated by the strong interconnected community structure often present in high-dimensional biomedical data. The UB structure partitions the covariance matrix into submatrices such that the covariance matrix can be represented using a block Hadamard product. We need more notation to describe this representation. First, let $\mathbf{A} = \text{diag}(a_1,\ldots,a_K)$, where $a_1,\ldots,a_K>0$, and  let $\mathbf{B} = (b_{ij})$ be a $K\times K$ covariance matrix. Let $\text{Id}_u$ be the $u\times u$ identity matrix, 
%$\text{J}_u$ be the $u \times u$ matrix of 1s, 
$\mathbf{1}_{u\times v}$ be the $u \times v$ matrix of 1s, and let $\mathbf{k} = (k_1,\ldots,k_K)$ be a vector of positive integers that indicate the size of each group, assuming $k_i > 1$ for each group $i=1,\ldots,K$. Define $\text{\bf{Id}}(\mathbf{k}) = \text{Bdiag}(\text{Id}_{k_1},\ldots,\text{Id}_{k_K})$, where $\text{Bdiag}(\cdot)$ constructs a block-diagonal matrix,  define $\mathbf{J}(\mathbf{k}) = (\mathbf{1}_{k_u \times k_v})$, and let $\circ$ be the block Hadamard product such that $\mathbf{A} \circ \mathbf{Id}(\mathbf{k}) = \text{Bdiag}(a_{1}\text{Id}_{k_1},\ldots,a_{K}\text{Id}_{k_K})$ and $\mathbf{B} \circ \mathbf{J}(\mathbf{k}) = (b_{uv}\mathbf{1}_{u \times v})$. With these notations and definitions in place, then $\Sigma$ has the UB structure if
\begin{align}\label{cov:UB0}
\Sigma = \mathbf{A} \circ \text{\bf{Id}}(\mathbf{k}) + \mathbf{B} \circ \mathbf{J}(\mathbf{k}).
\end{align}
Relating this to the covariance matrices of the random effects, note that the UB structure imposes further assumptions on $\Sigma_\epsilon$. Specifically, suppose we have within-group homogeneity, i.e., $\sigma^2_{\epsilon,i1} = \ldots = \sigma^2_{\epsilon,ik_i}$ for each $i = 1,\ldots, K$, and so we drop the last index to simply $\sigma^2_{\epsilon,i}$. This reduces the number of parameters of $\Sigma_\epsilon$ from $\sum_i k_i$ down to $K$, and we can instead use the representation $\widetilde{\Sigma}_\epsilon = \text{diag}(\sigma^2_{\epsilon,1},\ldots,\sigma^2_{\epsilon,K})$. From here, the linear mixed effects covariance model with the UB structure has the following form for the covariance matrix of the data:
\begin{align}\label{cov:UB}
\Sigma=  \widetilde{\Sigma}_\epsilon \circ \text{\bf{Id}}(\mathbf{k}) + \Sigma_\mu \circ \mathbf{J}(\mathbf{k}).
\end{align}

The within-group homogeneity imposed by the UB structure may not be desirable, and one can instead keep $\Sigma_\epsilon$ to be a diagonal matrix with no further restrictions on the diagonal entries. This leads to the linear mixed effects covariance model with the Heterogeneous-Block (HB) structure, which has the form
\begin{align}\label{cov:HB}
\Sigma =  \Sigma_\epsilon + \Sigma_\mu \circ \mathbf{J}(\mathbf{k}).
\end{align}
The HB structure still models the interconnected community structure like the UB structure, but allows for within-group heterogeneity. An important consideration, however, is the number of parameters. Note that $\widetilde{\Sigma}_\epsilon \circ \text{\bf{Id}}(\mathbf{k})$ has $K$ parameters, whereas $\Sigma_\epsilon$ has $\sum_i k_i$ parameters. Thus, the HB structure may be prohibitive for high-dimensional time series, especially if the length of the time series is short relative to the number of dimensions. Altogether, the choice of the structure depends on the data, but can also be determined by, e.g., exploratory data analyses on the diagonal entries within each group. 

We now describe how to estimate the parameters of the model for both the UB and HB structures. We arrange the time series data column-wise, keeping together the series that are in a group. Let $\mathbf{S}$ be the $\sum_i k_i \times \sum_i k_i$ sample covariance matrix for $\mathbf{y}$. Our goal is to use the group structure in the data to provide an estimated matrix that has either a UB or HB structure. Note that both a covariance matrix with either a UB or HB structure can be formed by block matrices with sizes determined by $\mathbf{k}$. We use $\mathbf{k}$ in the same way to partition $\mathbf{S}$ to blocks so that $\mathbf{S} = (\mathbf{S}_{uv})$. If we use the UB structure, \citet{yang24} derived the following method-of-moment estimates:
\begin{align}
    \widehat{\sigma}_{\mu,uv} &= 
        \begin{cases}
            \frac{\text{sum}(\mathbf{S}_{uv})}{k_uk_v}, & u \neq v, \\
            \frac{\text{sum}(\mathbf{S}_{uu}) - \text{tr}(\mathbf{S}_{uu})}{k_u(k_u-1)}, & u = v,
        \end{cases} \label{est:sig_mu}\\
    \widehat{\sigma}^2_{\epsilon,u} &= \frac{\text{tr}(\mathbf{S}_{uu})}{k_u} - \widehat{\sigma}_{\mu,uu} \label{est:sig_eps},
\end{align}
for $u,v = 1,\ldots,K$, where $\text{sum}(\cdot)$ denotes the sum of all the entries of the matrix, and $\text{tr}(\cdot)$ denotes the trace of the matrix. \citet{yang24} derived theoretical properties of the estimates in Equations \eqref{est:sig_mu} and \eqref{est:sig_eps}, which included strong consistency, asymptotic efficiency, and asymptotic normality.

Note that Equation \eqref{est:sig_mu} is the average of the diagonal elements (when $u = v$) or the average of the off-diagonal elements (when $u \neq v$) of $\mathbf{S}_{uv}$, and from Equation \eqref{est:sig_eps} we see that $\widehat{\sigma}_{\mu,uu} + \widehat{\sigma}^2_{\epsilon,u}$ is the average of all the diagonal elements of $\mathbf{S}_{uu}$. We use this same logic if we instead use the HB structure, where we only need to revise the estimate of $\Sigma_\epsilon$. Let $S_{ij}$ be the $(i,j)$-th element of $\mathbf{S}$. Then our revised estimate for the HB structure is
\begin{align}\label{est:sig_eps2}
    \widehat{\sigma}^2_{\epsilon,uj} &= \text{diag}(\mathbf{S}_{uu})_j - \widehat{\sigma}_{\mu,uu},
\end{align}
where $\text{diag}(\mathbf{S}_{uu})_j$ is the $j$-th diagonal element of $\mathbf{S}_{uu}$.

\subsection{Changepoint Estimation}\label{sec:pelt}
Within the mixed effect models described in the previous subsections, the parameters are fit via maximum likelihood, least squares, or similar methods.  Bringing back our reliance on time, the previous sections seek to optimise the no changepoint scenario
\begin{align}
    \sum_{s=1}^n \mathcal{C}(y_s|\theta)
    \label{eqn:optimind}
\end{align}
for $\theta$.  Here $\mathcal{C}(\cdot)$ is a measure of fit given fitted parameters $\hat{\theta}$ which, throughout this paper, will be twice the negative log-likelihood.  Recall that in describing the models for the individual segments we drop the subscript on $t$, here we similarly drop the subscripts $i,j$ and assume that 
\begin{align*}
    \mathcal{C}(y_s|\theta)=\sum_{i,j} \mathcal{C}(y_{s,i,j}|\theta_{i}).
\end{align*}
As written, \eqref{eqn:optimind} has a single model fit (and parameters) for all time points $s=1,\ldots,n$.  Recall that we define changepoints at times $0=\tau_0, \tau_1, \ldots, \tau_M, \tau_{M+1}=n$. Under the changepoint assumption the model parameters are restricted to be the same across segments of data and we seek to optimise,
\begin{align}
    \min_{\mathbf{\tau},M} \sum_{m=0}^M\sum_{s=\tau_m+1}^{\tau_{m+1}} \mathcal{C}(y_s|\hat{\theta}_{m+1}). \label{eqn:optimfree}
\end{align}
Due to the discrete nature of both the number and location of changepoints, standard estimation methods cannot be directly applied.  We now have a model selection problem where you need to select the appropriate number of changepoints and in turn, their location. This is akin to choosing the number of regressors in a regression problem.  Left free, the optimisation of \eqref{eqn:optimfree} would choose the maximum number of changepoints and so, as in the regression context, we need to penalize to obtain a parsimonious fit. \cite{PELTconsistency} demonstrates that penalties of the form $CM\log(n)$ are consistent for likelihood-based cost functions, $\mathcal{C}(\cdot)$, and constant $C$ with respect to $n$.  Thus we optimize,
\begin{align}
    F(n)=\min_{\mathbf{\tau},M} \sum_{m=0}^M\sum_{s=\tau_m+1}^{\tau_{m+1}} \mathcal{C}(y_s|\hat{\theta}_m) + CM\log(n). \label{eqn:fn}
\end{align}
Optimizing \eqref{eqn:fn} over all possible combinations of $M$ and $\mathbf{\tau}$ is a computationally intensive task. \cite{PELT} demonstrates how a combination of dynamic programming and pruning the search space can reduce the computational burden from $\mathcal{O}(2^n)$ to $\mathcal{O}(n)$.  Due to the independence of the segments, dynamic programming allows us to rewrite the search for all changepoints in \eqref{eqn:fn} in terms of the search for the last changepoint,
\begin{align}
    F(n)=\min_{\tau^*} F(\tau^*) + \sum_{s=\tau^*}^n \mathcal{C}(y_s|\hat{\theta}_M) + C\log(n). \label{eqn:dp}
\end{align}
Computing $F(\tau^*)$ recursively for $\tau^*=1,\ldots,n$ recovers the optimal set of changepoints for penalty $CM\log(n)$ in $\mathcal{O}(n^2)$ computational time.  To reduce this to $\mathcal{O}(n)$ one can prune the minimisation in \eqref{eqn:dp}.  As the minimisation is looking for the best last changepoint location, where there is an obvious changepoint location, the best last changepoint is unlikely to be prior to the obvious changepoint.  This intuition leads to a mathematically optimal pruning rule for an individual $\tau^*$, if it satisfies
\begin{align*}
    F(\tau^*)+\sum_{s=\tau^*}^t \mathcal{C}(y_s|\hat{\theta}_M) \geq F(t). 
\end{align*}
Intuitively this says that if at any time in the recursive computation, a candidate last changepoint location, $\tau^*$, is more than $C\log(n)$ larger than the optimal likelihood at that time, $t$, it can never be the last changepoint at any future point so it can be pruned from the minimisation in \eqref{eqn:dp}. It was shown in \cite{PELT} that this pruning rule exactly solves the same minimisation problem without approximation.  The authors call this algorithm, PELT, Pruned Exact Linear Time.

We use PELT as a wrapper for each of the mixed effect model forms discussed earlier in this section, by using the appropriate negative twice the log-likelihood as $\mathcal{C}(\cdot)$ in \eqref{eqn:dp}.  The computational cost of this is then $\mathcal{O}(Ln)$ where $L$ is the computational order of evaluating the likelihood for a single segment.  If summary statistics can be utilised as is possible in \eqref{eqn:glme} then $L=p$, the number of regressors.

\section{Simulation Study}\label{sec:sim}
In this section we will assess the different base random effects models via simulation studies inspired by our applications in digital health.  We denote the different methods as, PELT-GLMER for the generalised regresion model, PELT-LMEC-UB for the covariance model assuming uniform blocks, and PELT-LMEC-HB for the covariance model assuming heterogeneous blocks.  Our simulations intend to show when the approaches perform well alongside violations of assumptions that may occur in applications which will deteriorate performance.

% Performance metrics
We assess performance using the following metrics: number of detected changepoints ($|M - \hat{M}|$), mean absolute error (MAE), defined as 
\begin{align*}
\text{MAE} = \frac{1}{n}\sum_{t=1}^n||\widehat{\theta}_t - \theta_t||_1,
\end{align*}
where $\theta$ is the set of estimated parameters, e.g. binomial probability or covariance matrix.  The MAE assesses the impact of the estimated changepoints on the estimates of the parameters across all time points and should be viewed in conjunction with the number and location of changepoints.  For some applications accurately estimating the number and location of changepoints is the primary objective, for others, the parameter estimation is paramount and the changepoints are a nuisance.
% Elaborating on the first three metrics, for each simulation setting we compare the set of estimated changepoint locations $\widehat{\mathbf{\tau}} = \{\hat{\tau}_1,\ldots,\hat{\tau}_{\hat{M}}\}$ with the true set of changepoint locations $\mathbf{\tau} = \{\tau_1,\ldots,\tau_M\}.$ Note that $M=3$ in the multiple changepoint simulation setting, and $M=1$ in the other simulation settings. We say that the true changepoint $\tau_i$ was correctly detected if, for some tolerance window $h$, $|\tau_i - \hat{\tau}_j| \leq h$ for some $1\leq j \leq \hat{M}$. Let $\mathbf{\tau}_c$ be the set of correctly estimated changepoints. Then we calculate $\text{TPR} = |\mathbf{\tau}_c|/\mathbf{\tau}$ and $\text{FPR} = |\mathbf{\tau}_c - \mathbf{\tau}|/\mathbf{\tau}_c$. 

\subsection{Practical Considerations}\label{sec:practical}
Our proposed method has three practical concerns: i) choice of parameters, ii) the grouping structure, and iii) model assumptions. 

% Choice of parameters - minseglen and penalty
The minimum segment length balances the need for having enough data to estimate the model parameters (regression or covariance matrix) with being localised enough in time. Consequently, our method may miss changepoints that occur rapidly. For the PELT-GLMER model, we a minimum segmet length of $p$.  For the covariance model, \citet{ryan23} used a minimum segment length of $4p$, but from our explorations of the impact of this parameter on the performance of the method, we recommend a smaller value of $2p$ for the minimum segment length. Note that PELT-LMEC-UB has $K+{{K+1}\choose 2}$ parameters to estimate whereas PELT-LMEC-HB has $p+{{K+1}\choose 2}$ parameters. Assuming that $K$ is much smaller than $p$, the value of $2p$ for the minimum segment length provides a sufficient amount of data for estimating the parameters for either method.

Recall the penalty term in PELT, $CM\log(n)$. Another parameter to set is the constant $C$ in this penalty term. If $C$ is too small, then there will be more candidate changepoints, thereby increasing the possibility for falsely detected changepoint locations. If $C$ is too large, then only changepoints that lead to very large changes in the likelihood will be detected. For all methods we use a BIC-like penalty; $C=p$ for PELT-GLMER and $C = K+ {{K+1}\choose 2}$ for PELT-LMEC-UB and PELT-LMEC-HB, which is also the number of parameters to estimate in the covariance matrix for LMEC-UB. This leads to a BIC-like penalty term. In investigations not shown here, setting $C=p+{{K+1} \choose 2}$, which is the number of parameters for LMEC-HB, led to unsatisfactory performance for changepoint detection, especially whenever $p$ was relatively large.

% Grouping structure
Our proposed method assumes a fixed and known grouping structure for the time series data. This affects the estimates of the blocks in each of the UB and HB structures, thereby affecting model fit. One could use a community detection or clustering algorithm to identify the groups in a data-driven manner \citep{Wu21}. We show in our simulation study the impact of misspecified groups on the performance. In our data example, we used domain knowledge about the study and the data to create the groups.

% Model assumptions
Our proposed method assumes no temporal autocorrelation, which is often present in time series. Prior works showed that temporal autocorrelation leads to an overestimation on the number of changepoints \citep{Shi22,gallagher22,ryan23}.  \citet{gallagher22} recently showed how one can incorporate the autocorrelation structure of the data for changepoint detection in univariate time series, but adapting this to multivariate time series is beyond the scope of this work. In our simulations, we show the detrimental effects of autocorrelation on our proposed method, and in our data example we pre-whitened the time series.
    
\subsection{Generalised Linear Mixed Effects Model}
Motivated by different approaches to our application we consider Bernoulli data from $K=4$ groups (people), each time series represents either a day ($n=96$) or a week ($n=672$), and we have either 14 or 56 replicates (2/6 weeks of days and approximately a quarter/year for weekly data).  The global mean probability can either be constant or vary at changepoints, each replicate will independently simulate these from a Uniform (0.1,0.9) distribution.  Similarly, the group effect can either be constant or vary at changepoints with the added scenario where a single group (final) or all groups can change at each changepoint.  The group effect is independently simulated from a Normal distribution with mean 0 and standard deviation 0.05.  When simulating the data, we curtail the sum of the group and global effect to be between 0.01 and 0.99.  The smaller daily series are simulated with three changepoints at 28, 34 and 88. The larger weekly series are simulated with 14 changepoints at 27, 87, 121, 181, 222, 282, 316, 376, 410, 470, 511, 571, 605, 665 representing wake and sleep with some variability across days. For each scenario, we simulate 500 replicates.

% RK to add PELT-GLMER results here when finished on the cluster.

\subsection{Linear Mixed Effects Covariance Model}
We used five  scenarios for our simulation study: 
\begin{enumerate}
    \item[1.] \textbf{Zero changepoints}. We simulated data from a zero-mean multivariate normal distribution with random covariance matrices. This scenario assesses the false positive rate for each method.
    \item[2.] \textbf{One changepoint}. We set the changepoint at $\tau_1 = n/2$, and within each segment, we simulated data from a zero-mean multivariate normal distribution with random covariance matrices.
    \item[3.] \textbf{Multiple changepoints}. We set the changepoints at $\tau_1 = 0.4n, \tau_2=0.6n$, and $\tau_3=0.75n$. Within each segment, we similarly simulated data from a zero-mean multivariate normal distribution with random covariance matrices.
    \item[4.] \textbf{Multiple changepoints, misspecified groups}. We follow the same setup as in Scenario 3. However, we deliberately misspecified too few groups to assess the impact of group misspecification.
    \item[5.] \textbf{Multiple changepoints, autocorrelated data}. We follow the same setup as in the Scenario 3. However, within each segment, we set $y_{i,j,t} = \phi y_{i,j,t-1} + z_{i,j,t}$, where, for a given $t$, we simulated $\mathbf{z}_{t}$ from a multivariate normal distribution with random covariance matrices. Note that our method assumes temporal independence, thus, this scenario assesses of the impact of temporal autocorrelation.
\end{enumerate}
In each of the above scenarios, we considered $n \in \{500, 1000\}$ and $K=4$ groups such that all groups contain $k_i=5$ or $k_i=15$ many time series. Thus, the dimensions for both $\Sigma$ and $\Sigma_\epsilon$ was either $20\times 20$ or $60\times 60$. For Scenario 4, the  dimensions for $\Sigma_\mu$ was $5\times 5$, but for the other scenarios its dimensions were $4 \times 4$. Note for Scenario 4 that this implies that there were truly 5 groups, but in the model we set $K=4$. To construct $\Sigma_\mu$, we first simulated a random matrix from the Wishart distribution with degrees of freedom set to the number of dimensions of the data (i.e., either 20 or 60) with scale matrix the identity matrix of appropriate dimension, and then converted the simulated matrix to a correlation matrix. To construct $\Sigma_\epsilon$ we created a diagonal matrix with each diagonal entry simulated from the Uniform$(0.75,1.25)$ distribution. For scenario 5 with autocorrelated data, we used the above for simulating a structured covariance matrix to create the covariance matrix for $\mathbf{z}_t$, and we set $\phi=0.6$. 

% Describe ratio method and BSCOV
We compare the performance of each of PELT-LMEC-UB and PELT-LMEC-HB in each of the above scenarios to two competing methods: the Ratio method \citep{ryan23} and the Wild Sparsified Binary Segmentation (BSCOV) \citep{Li23}, which we now briefly describe. The Ratio method uses a test statistic based on the eigenvalues of the covariance matrices calculated between two segments of the data. The Ratio method uses binary segmentation for changepoint detection. BSCOV, on the other hand, fits an approximate factor model to the multivariate time series data, and then uses wild binary segmentation, developed by \citet{Fryzlewicz14}, for identifying the changepoints in the factor components.

%%%%% Describe results
% No changepoints
In Tables \ref{tab:lmec} and \ref{tab:lmec2}, we see how each method performs in estimating the number of changepoints and the covariance matrices. In the case where there was truly no changepoints present, all methods performed very well with respect to their false positives. For the MAE in Table \ref{tab:lmec2}, we see that PELT-LMEC-HB and PELT-LMEC-UB both performed better than Ratio, and this is due to PELT-LMEC-HB and PELT-LMEC-UB assuming a structure for the covariance matrix whereas Ratio was unstructured. Interestingly, even though the true covariance matrices was heterogeneous along the diagonal entries, assuming a UB structure led to better performance with respect to MAE even at the higher dimension, and this is likely due to the lower variance in the estimates as a result of averaging the estimates of the diagonal entries within a block.

\begin{figure}
    \centering
    \includegraphics[width=1\linewidth]{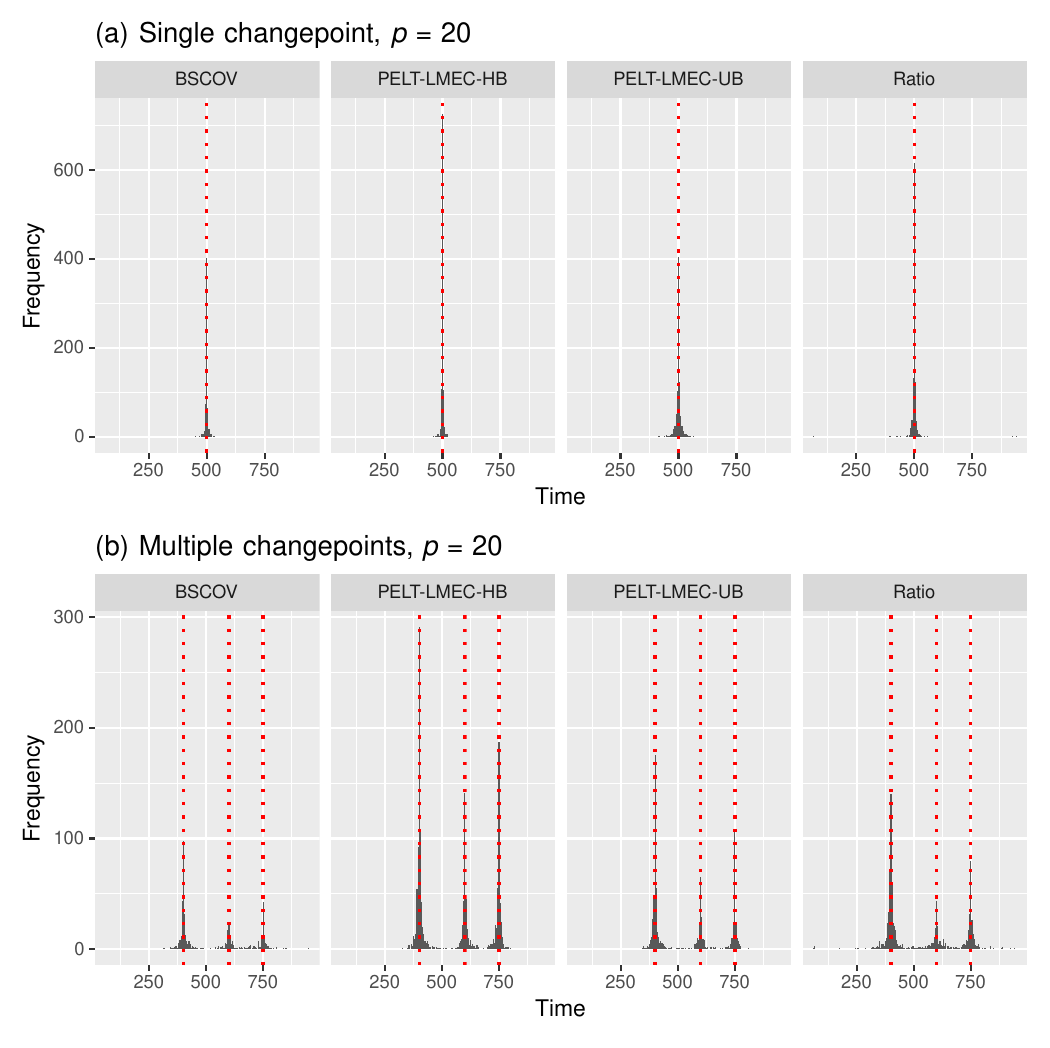}
    \caption{Histogram of the location of the detected changepoints in the simulations from two scenarios obtained from each of the four methods. Dotted red lines correspond to the  true changepoint locations. (a) Single changepoint,w ith $p=20$ and $n=1000$. (b) Multiple changepoints, with $p=20$ and $n=1000$.}
    \label{fig:lmec1}
\end{figure}

% Single changepoint
When there was a single changepoint, PELT-LMEC-HB performed the best at estimating only one  changepoint. The Ratio method also performed very well, and BSCOV performed the worst of the four methods. We also see that as the sample size increased each method performed better at estimating the number of changepoints. We also see that the MAEs were similar or better was the dimensionality increased. From Figure \ref{fig:lmec1}(a), we see that, for $p=20$ and $n=1000$, the location of the changepoints, if a method detected one, was at or very close to the true changepoint location.

\begin{figure}
    \centering
    \includegraphics[width=1\linewidth]{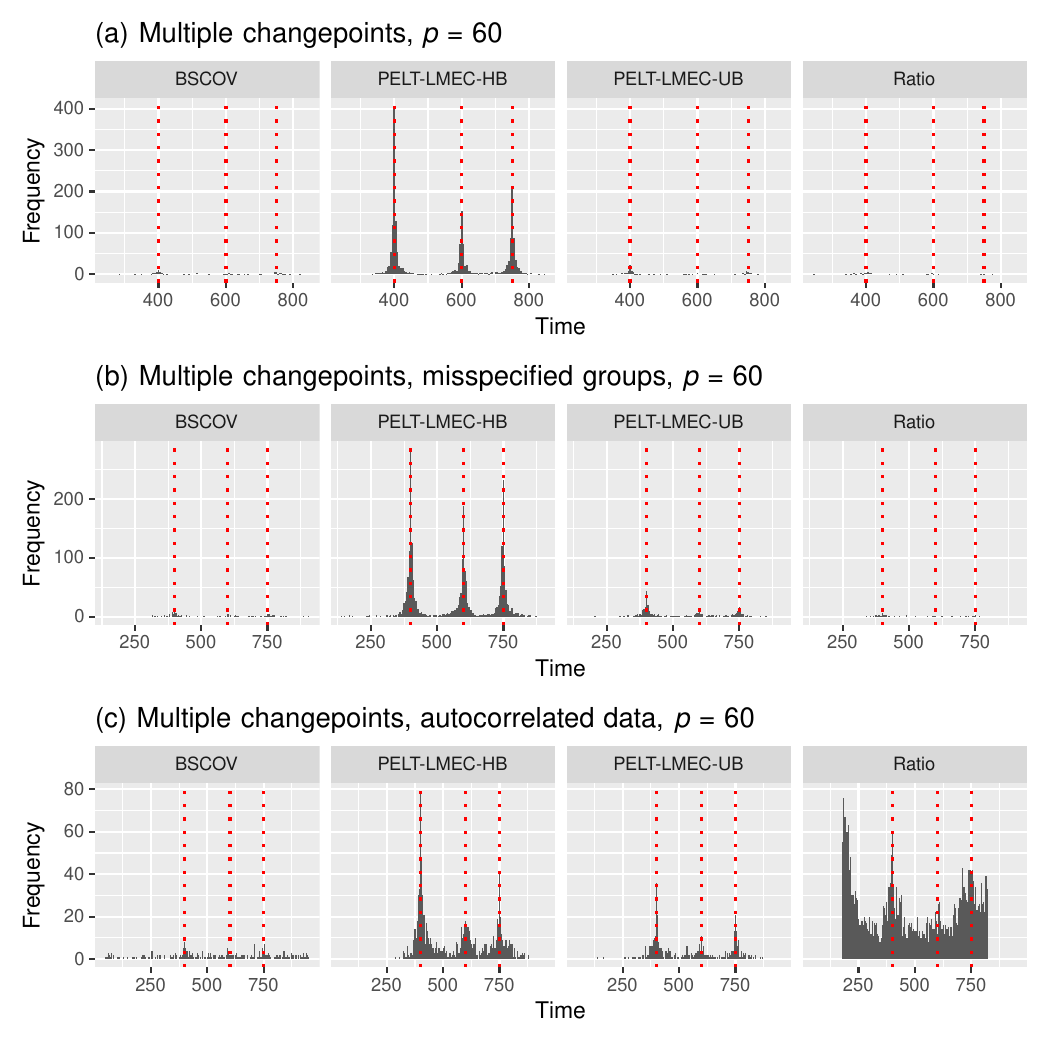}
    \caption{Histogram of the location of the detected changepoints in the simulations from two scenarios obtained from each of the four methods. Dotted red lines correspond to the  true changepoint locations. (a) Multiple changepoints, with $p=60$ and $n=1000$. (b) Multiple changepoints and misspecified groups, with $p=60$ and $n=1000$. (c) Multiple changepoints and autocorrelated data, with $p=60$ and $n=1000$.}
    \label{fig:lmec2}
\end{figure}

 % Multiple changepoints
When there were $M=3$ changepoints, all methods struggled at detecting all three changepoints, especially when $p=60$. Recall that $\tau_2 = 0.6n$ and $\tau_3=0.75n$, and so since they were relatively close in time it was difficult for the methods to detect both of these changepoints. Indeed, from Figure \ref{fig:lmec1}(b) we see that $\tau_1$ was the easiest changepoint for all methods to detect since that changepoint had the greatest frequency of being detected relative to the other two changepoints. However, when $p=60$ and $n=500$, all methods struggled to detect any of the changepoints. For $p=60$ and $n=1000$, PELT-LMEC-HB performed the best since it usually detected at least two of the changepoints, with $\tau_2$ having the lowest frequency of being detected as shown in Figure \ref{fig:lmec2}(b). However, the other methods continued to struggle to detect any of the changepoints. Indeed, from Figure \ref{fig:lmec2}(a), we see the poor performances of all methods except PELT-LMEC-HB at detecting the changepoints. Note, however, that PELT-LMEC-UB generally had lower MAEs than PELT-LMEC-HB as shown in Table \ref{tab:lmec2}, so even though the uniform assumption led to better estimates of the covariance matrices, the heterogeneity assumption led to better estimates of the number and location of changepoints.

% Group misspecification
When the number of groups specified for either PELT-LMEC-HB or PELT-LMEC-UB was fewer than the true number of groups, their respective MAEs were smaller than the scenario when $M=3$ but with correctly specified groups, as shown in Table \ref{tab:lmec}. This is likely due to the greater amount of averaging across the dimensions of the time series, leading to a decrease in the variance of the estimates that overcame the increase in the bias. While it may also seem that both PELT-LMEC-HB and PELT-LMEC-UB better detected the number of changepoints, we point out that, from a visual assessment of Figure \ref{fig:lmec2}, the precision of the estimates of the changepoint locations appear to be lower, since the distribution of the estimated changepoints appear to be wider around the true changepoint locations.

% Autocorrelation
Autocorrelation generally inflates false positives in changepoint detection methods that were not explicitly designed to account for autocorrelation \citep{gallagher22,Shi22,ryan23}, and we observe that phenomenon here. In Figure \ref{fig:lmec2}(b), we see that all methods are not as precise at estimating the changepoint locations. While Table \ref{tab:lmec} suggests that the Ratio method can estimate very well the number of changepoints for $p=60$ and $n=1000$, we see in Figure \ref{fig:lmec2}(b) that the location of these changepoints are not temporally precise at the true changepoint locations, with the most common changepoint it detected not anywhere close to the location of a true changepoint. PELT-LMEC-HB and PELT-LMEC-UB both better estimate the changepoint locations relative to BSCOV and Ratio, but these two methods also yield imprecise estimates of the changepoint locations.

% \begin{ltable}
\begin{table}
\centering
\begin{tabular}{lrrrrrr}
& & & \multicolumn{4}{c}{$\vert M - \hat{M}\vert$} \\
\cmidrule(lr){4-7}
Simulation & & & & PELT & PELT & \\
      Setting & $p$ & $n$ & BSCOV & LMEC-HB & LMEC-UB & Ratio  \\ \hline
  \hline
$M=0$  & 20 & 500 & \textbf{0.00} & \textbf{0.00} & \textbf{0.00} & 0.02  \\ 
    &  & 1000 & \textbf{0.00} & \textbf{0.00} & \textbf{0.00} & \textbf{0.00}   \\ 
   & 60 & 500 & \textbf{0.00} & \textbf{0.00} & \textbf{0.00} & \textbf{0.00}   \\ 
   &  & 1000 & \textbf{0.00} & \textbf{0.00} & \textbf{0.00} & \textbf{0.00}  \\ 
  $M=1$  & 20 & 500 & 0.68 & \textbf{0.00} & 0.50 & 0.04  \\ 
    &  & 1000 & 0.40 & \textbf{0.00} & 0.15 & 0.01  \\ 
   & 60 & 500 & 0.93 & \textbf{0.00} & 0.69 & 0.04    \\ 
   &  & 1000 & 0.89 & \textbf{0.00} & 0.31 & \textbf{0.00}  \\ 
  $M=3$  & 20 & 500 & 2.76 & \textbf{2.53} & 2.80 & 2.72 \\ 
    &  & 1000 & 2.51 & \textbf{1.54} & 2.20 & 2.08   \\ 
   & 60 & 500 & 2.93 & \textbf{2.45} & 3.00 & 3.00  \\ 
   &  & 1000 & 2.93 & \textbf{1.06} & 2.90 & 2.96 \\ 
  $M=3$ & 20 & 500 & 2.65 & 2.55 & 2.81 & \textbf{2.52} \\ 
   misspecified    &  & 1000 & 2.31 & 1.71 & 2.24 & \textbf{1.54}\\ 
  groups  & 60 & 500 & 2.92 & \textbf{1.73} & 2.94 & 3.00   \\ 
    &  & 1000 & 2.90 & \textbf{0.60} & 2.69 & 2.95  \\ 
  $M=3$, & 20 & 500 & \textbf{1.96} & 2.33 & 2.44 & 2.16  \\ 
  autocorrelation   &  & 1000 & 2.37 & \textbf{1.45} & 1.73 & 8.38  \\ 
    & 60 & 500 & 2.68 & 2.81 & 2.87 & \textbf{2.00}\\ 
    &  & 1000 & 2.75 & 2.07 & 2.62 & \textbf{0.16}  \\ 
   \hline
\end{tabular}
\caption{$|M-\hat{M}|$ for the estimated covariance matrix across all scenarios. The method that performed the best at a given simulation setting and ($p,n$) combination had its entry in boldface font.}
\label{tab:lmec}
\end{table}
 %\end{ltable}
 
\begin{table}
\centering
\begin{tabular}{lrrrrr}
& & &\multicolumn{3}{c}{MAE} \\
\cmidrule(lr){4-6}
Simulation & & & PELT & PELT & \\
      Setting & $p$ & $n$ & LMEC-HB & LMEC-UB & Ratio \\ \hline
  \hline
$M=0$  & 20 & 500 & 443.67 & \textbf{419.37} & 800.86 \\ 
    &  & 1000 & 442.03 & \textbf{418.22} & 795.80 \\ 
   & 60 & 500   & 3079.58 & \textbf{2989.64} & 7049.65 \\ 
   &  & 1000  & 3008.23 & \textbf{2914.25} & 6986.55 \\ 
  $M=1$  & 20 & 500  & 465.24 & \textbf{440.36} & 816.96 \\ 
    &  & 1000  &  450.39 & \textbf{431.95} & 813.02 \\ 
   & 60 & 500  & \textbf{3071.74} & 3084.37 & 7140.78 \\ 
   &  & 1000  &  3111.78 & \textbf{3073.81} & 7093.82 \\ 
  $M=3$  & 20 & 500 &  379.61 & \textbf{366.79} & 555.48 \\ 
    &  & 1000  & 383.31 & \textbf{369.59} & 552.91 \\ 
   & 60 & 500 &  2914.51 & \textbf{2813.91} & 4823.67 \\ 
   &  & 1000  &   2971.55 & \textbf{2820.12} & 4859.30 \\ 
  $M=3$ & 20 & 500  &  291.98 & \textbf{276.67} & 558.65 \\ 
   misspecified    &  & 1000 &  292.60 & \textbf{276.30} & 556.11 \\ 
  groups  & 60 & 500  & 2093.32 & \textbf{2000.45} & 4787.28 \\ 
    &  & 1000  &  2136.93 & \textbf{2023.94} & 4853.28 \\ 
  $M=3$, & 20 & 500 &  547.18 & \textbf{521.74} & 827.61 \\ 
  autocorrelation   &  & 1000   & 543.32 & \textbf{528.35} & 826.35 \\ 
    & 60 & 500 &  3972.77 & \textbf{3779.38} & 6912.02 \\ 
    &  & 1000 &  4142.04 & \textbf{3936.08} & 7172.31 \\ 
   \hline
\end{tabular}
\caption{Mean absolute error (MAE) for the estimated covariance matrix across all scenarios. Since BSCOV fits a factor model to the data and does not explicitly estimate covariance matrices, we did not evaluate its MAE for estimating the covariance matrices, and hence was omitted. The method that performed the best at a given simulation setting and ($p,n$) combination had its entry in boldface font.}
\label{tab:lmec2}
\end{table}

\section{Applications}\label{sec:app}

\subsection{At-Home Digital Health Monitoring}
Traditionally, point of contact with healthcare professionals was the way clinicians could monitor someone's health clinically.  Over time, this leads to increased stays in hospital, inconvenient repeated appointments, or discharge followed by adverse events in order to monitor effectively.  With the advent of digital health monitoring, a patient can permit healthcare providers to receive updates from tested and approved devices at home.  Similarly, those interested in monitoring their long-term health can purchase devices that track health metrics over time.

Many of these devices are wrist-worn for continuous measurement or reminders are given through apps to take measurements.  These devices often have low adherence in specific patient groups.  An alternative approach is passive monitoring within the home.  These devices provide a different type of data which can often be combined with other health measurements when available.

We analyse data from four people who use the Howz at-home passive activity monitoring.  The passive sensors record movement or interaction with an object, e.g.\ opening the fridge, or turning the kettle on.  Howz summarize the data as a yes/no for whether there is any activity within a 15-minute window.  This results in 96 yes/no observations per day and we analyse 56 days of activity for each person.  

Utilising the method described in section \ref{sec:glme} we fit a Bernoulli distribution with logit link function to each day of data.  Each day of observations from an individual is a separate time series within the same group, thus we have four groups.  Prior analysis of similar Howz data \citep{HowzSimonRSSC,HowzJessSIM} has identified that individuals in this cohort tend to have similar behaviour across days.  Thus we expect to see global effects such as similar wake-up and bedtime changes in activity levels alongside group effects related to an individual's routine.  Being able to describe routines that are common across individuals as well as individual-level behaviour is important in providing insights for different purposes.

\begin{figure}
    \centering
    \includegraphics[width=1\linewidth]{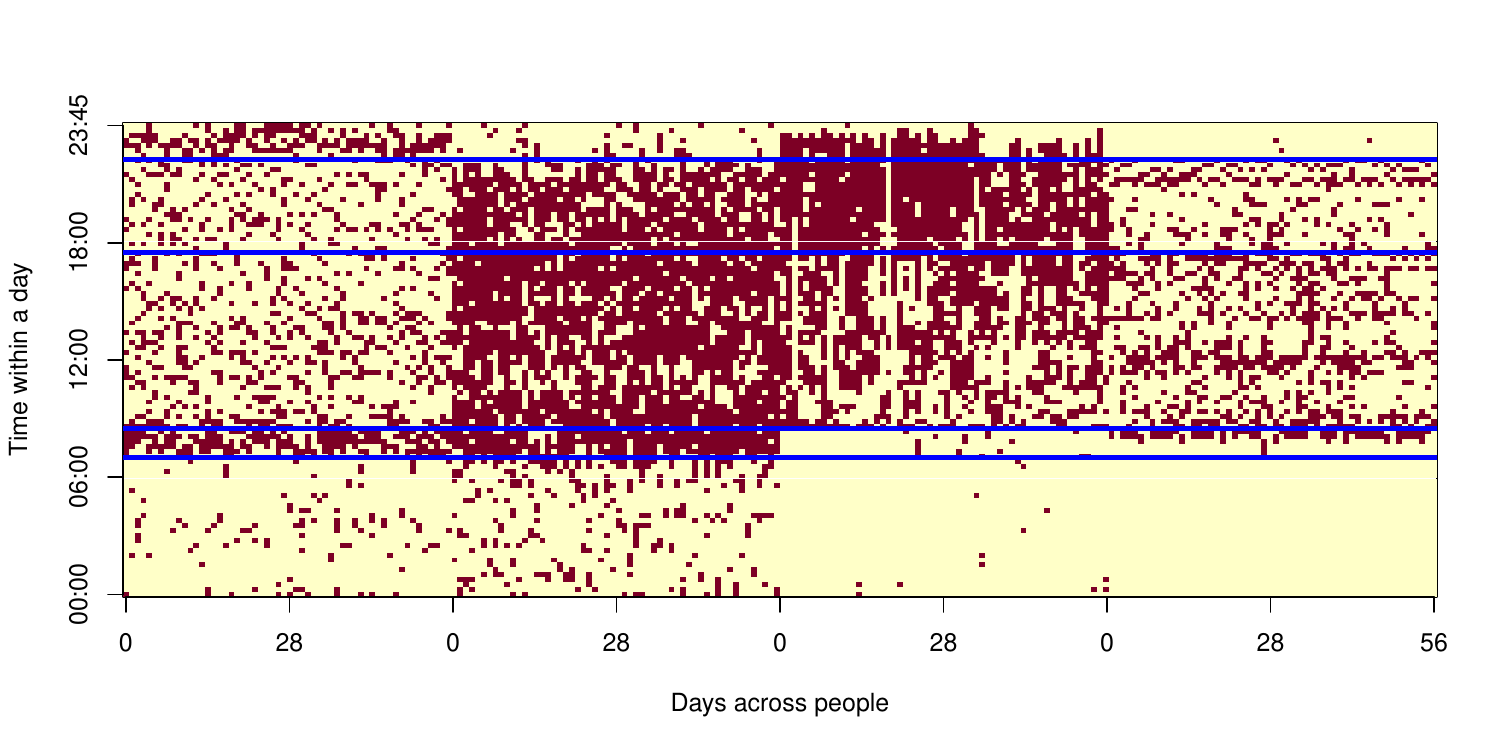}
    \caption{Daily Howz data with changepoints at 7am, 8:30am, 5:30pm, 10:15pm. The $x$-axis shows days for each person, across four people.}
    \label{fig:Howzdaily}
\end{figure}

\begin{table}[ht]
\centering
\begin{tabular}{rrrrrr}
  \hline
Segment & 1 & 2 & 3 & 4 & 5 \\ 
  \hline
2.5\% level & 0.000 & 0.109 & 0.278 & 0.229 & 0.024 \\ 
  Mean & 0.011 & 0.367 & 0.483 & 0.497 & 0.141 \\ 
  97.5\% level & 0.176 & 0.732 & 0.698 & 0.752 & 0.501 \\ 
   \hline
\end{tabular}
\caption{Fixed effects and 95\% bootstrap confidence intervals for the fixed effect probability of activity across segments in Howz daily analysis.}\label{tbl:HowzDaily}
\end{table}

\begin{table}[ht]
\centering
\begin{tabular}{rrrrr}
  \hline
Person & 1 & 2 & 3 & 4 \\ 
Segment &&&& \\  \hline
1 & 8.135 & 13.246 & -1.488 & -16.737 \\ 
  2 & 3.227 & 7.309 & -8.749 & -1.707 \\ 
  3 & -3.145 & 4.292 & 0.599 & -1.746 \\ 
  4 & -4.310 & 2.903 & 5.284 & -3.878 \\ 
  5 & 5.258 & -0.873 & 5.499 & -9.470 \\ 
   \hline
\end{tabular}
\caption{Contribution of the random effects, on the logit scale,  of activity across segments and people in Howz daily analysis.}\label{tbl:HowzDailyRandom}
\end{table}

We ran PELT-GLMER with a penalty of $56\log(96)$ for 56 days of data each with 96 observations.  Figure \ref{fig:Howzdaily} shows the data with the changepoints overlayed.  It is clear that the first changepoint (7am) is driven by a change in the first two participants.  The second change (8:30am) looks to be driven by participants 1, 3, and 4.  The 5:30pm change is driven by participant 3 increasing, and participant 4 decreasing activity.  The final change at 10:15pm looks to be bedtime for participants 2 and 4, and the start of a bedtime routine (increase in activity) for participant 1.  Table \ref{tbl:HowzDaily} depicts the fixed effect, (global mean) for each segment along with bootstrap 95\% confidence intervals.  This reflects the fact that the overall probability of activity is higher during the day than in the two nighttime periods, and increases as the day progresses.  Table \ref{tbl:HowzDailyRandom} then depicts the deviation, on the logit scale, of the groups (each person) from this average behaviour.  This correlates with what we can determine visually, showing the largest variation across individuals is in the initial nighttime (00:00-07:30) period corresponding to individuals who conduct activity (e.g., visiting the bathroom) over night frequently to not at all.

\begin{figure}
    \centering
    \includegraphics[width=1\linewidth]{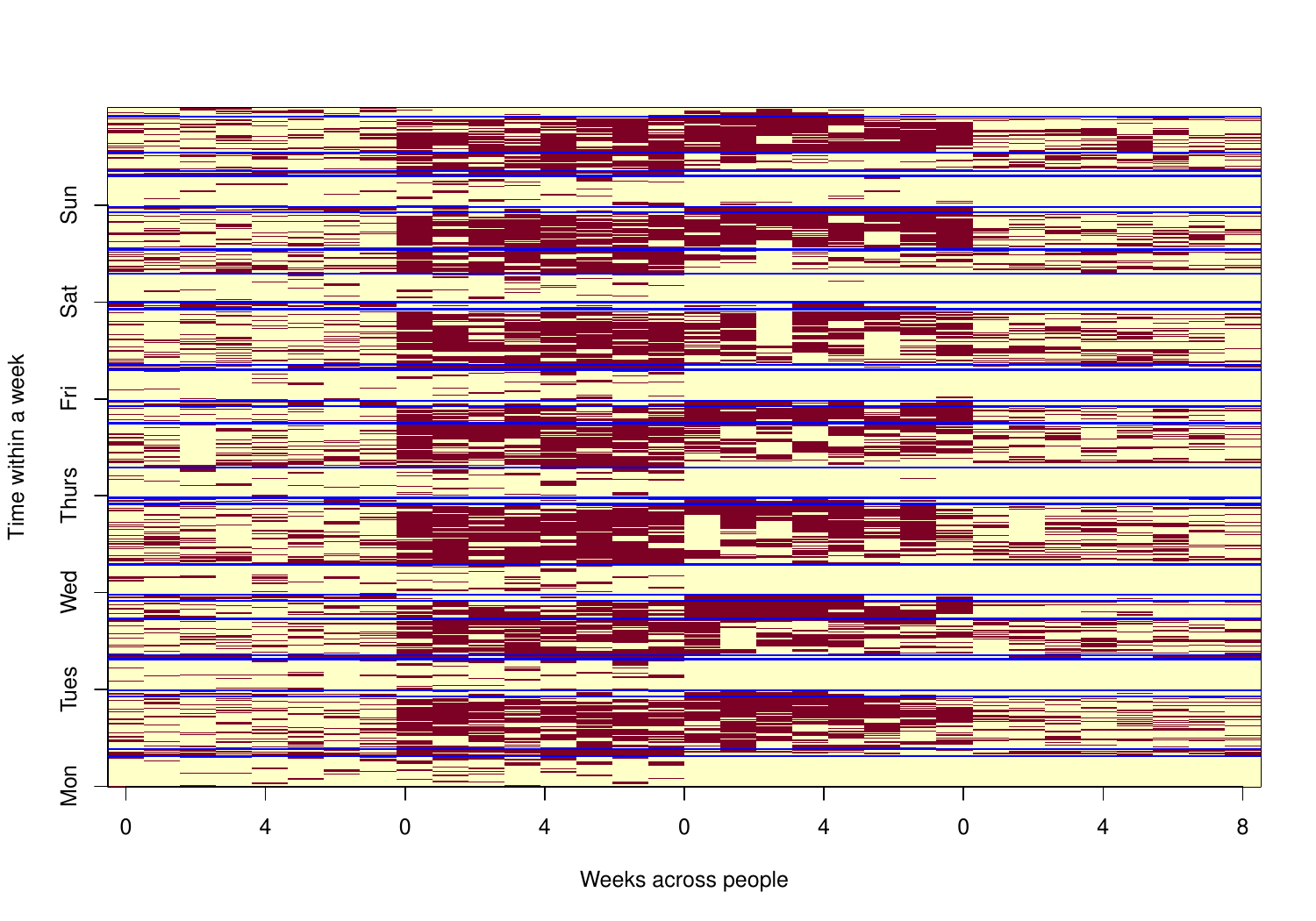}
    \caption{Weekly Howz data with changepoints. The $x$-axis shows weeks for each person, across four people.}
    \label{fig:Howzweekly}
\end{figure}

\begin{table}[ht]
\centering
\begin{tabular}{rrrrrrrrrrr}
  \hline
Segment & 1 & 2 & 3 & 4 & 5 & 6 & 7 & 8 & 9 & 10 \\
  \hline
2.5\% level & 0.000 & 0.349 & 0.238 & 0.062 & 0.000 & 0.108 & 0.257 & 0.182 & 0.044 & 0.000 \\
  Mean & 0.005 & 0.600 & 0.469 & 0.221 & 0.004 & 0.435 & 0.457 & 0.511 & 0.224 & 0.005 \\
  97.5\% level & 0.178 & 0.825 & 0.713 & 0.526 & 0.192 & 0.817 & 0.670 & 0.825 & 0.604 & 0.120 \\
   \hline
 Segment  & 11 & 12 & 13 & 14 & 15 & 16 & 17 & 18 & 19 & 20 \\
 2.5\% level & 0.272 & 0.057 & 0.000 & 0.245 & 0.220 & 0.016 & 0.000 & 0.027 & 0.261 & 0.012 \\
 Mean & 0.504 & 0.213 & 0.017 & 0.458 & 0.517 & 0.200 & 0.015 & 0.463 & 0.465 & 0.148 \\
 95.5\% level & 0.739 & 0.547 & 0.159 & 0.662 & 0.830 & 0.692 & 0.116 & 0.971 & 0.677 & 0.535 \\
 \hline
 Segment & 21 & 22 & 23 & 24 & 25 & 26 & 27 & 28 & 29 &\\ 
 2.5\% level & 0.002 & 0.254 & 0.274 & 0.000 & 0.004 & 0.000 & 0.239 & 0.292 & 0.008 &\\ 
 Mean & 0.019 & 0.457 & 0.533 & 0.266 & 0.034 & 0.158 & 0.442 & 0.554 & 0.086 &\\ 
 97.5\% level & 0.110 & 0.675 & 0.795 & 0.874 & 0.154 & 0.887 & 0.663 & 0.802 & 0.392 &\\ 
\end{tabular}
\caption{Fixed effects and 95\% bootstrap confidence intervals for the fixed effect probability of activity across segments in Howz weekly analysis.}\label{tbl:HowzWeekly}
\end{table}

\begin{table}[ht]
\centering
\begin{tabular}{rrrrr}
  \hline
Person & 1 & 2 & 3 & 4 \\ 
 Segment &&&& \\  \hline
1 & 3.286 & 4.019 & -2.065 & -2.065 \\ 
  2 & 0.097 & 1.491 & -1.311 & -0.316 \\ 
  3 & -1.080 & 0.992 & 1.003 & -0.913 \\ 
  4 & 0.883 & -0.192 & 1.366 & -1.871 \\ 
  5 & 3.201 & 4.041 & -2.051 & -2.051 \\ 
  6 & 1.631 & 1.297 & -2.570 & -0.239 \\ 
  7 & -1.051 & 1.222 & 0.251 & -0.419 \\ 
  8 & -1.122 & 0.641 & 2.041 & -1.568 \\ 
  9 & 1.369 & -1.306 & 1.786 & -1.620 \\ 
  10 & 2.952 & 3.775 & -1.964 & -1.964 \\ 
  11 & -0.905 & 1.511 & 0.199 & -0.807 \\ 
  12 & 1.008 & -1.195 & 1.570 & -1.195 \\ 
  13 & 1.666 & 2.758 & -1.205 & -2.310 \\ 
  14 & -0.757 & 1.461 & -0.026 & -0.676 \\ 
  15 & -1.621 & 0.728 & 2.029 & -1.145 \\ 
  16 & 1.457 & -0.334 & 2.181 & -2.822 \\ 
  17 & 1.753 & 2.458 & -1.154 & -2.207 \\ 
  18 & 0.443 & 3.389 & -3.123 & -0.688 \\ 
  19 & -1.106 & 1.268 & 0.354 & -0.514 \\ 
  20 & 1.864 & -0.184 & 1.441 & -2.676 \\ 
  21 & 1.116 & 2.152 & -0.682 & -2.004 \\ 
  22 & -0.328 & 1.515 & -0.791 & -0.393 \\ 
  23 & -1.322 & 1.083 & 1.200 & -0.963 \\ 
  24 & 1.104 & -0.362 & 3.392 & -3.525 \\ 
  25 & 0.594 & 1.773 & 0.362 & -2.358 \\ 
  26 & 2.614 & 3.785 & -3.308 & -1.846 \\ 
  27 & -0.726 & 1.381 & -0.794 & 0.145 \\ 
  28 & -1.226 & 0.743 & 1.363 & -0.885 \\ 
  29 & 1.612 & -0.595 & 1.750 & -2.254 \\ 
   \hline
\end{tabular}
\caption{Contribution of the random effects, on the logit scale, of activity across segments and people in Howz weekly analysis}\label{tbl:HowzWeeklyRandom}
\end{table}

When we consider each day as a time series within the same group we are implicitly assuming that each day of the week behaves similarly.  This may not be the case for individuals and prior analysis of this data in \cite{Taylor21} indicated there could be day of the week effects in some individuals.  To assess this, we conducted the same analysis but creating each time series from a week of data instead of a day.  The method is agnostic to this choice and we simply modify the penalty to be $8\log(96*7)$ to reflect the dimension change.

Figure \ref{fig:Howzweekly} depicts the weekly changepoints there are at least three, usually four, changepoints each day.  The changepoints in the weekly analysis are very similar to the daily analysis.  The first changepoint (07:30am) is identified either exactly (3 days) or within 30 mins (4 days).  Similarly, the final changepoint (10:15pm) is identified eactly (4 days) or within 15 mins (3 days).  The changepoint at 5:30pm is only identified exactly on Tuesday and at 6:00pm on Thursday.  Tables \ref{tbl:HowzWeekly} and \ref{tbl:HowzWeeklyRandom} give the fixed effects and random contribution to the probability of events occurring in each segment.  These show that whilst there is some variability in the changepoints across days, the morning wake up and evening wind down times are fairly consistent across days of the week, the variability is in the probability of activity within a day.  These findings are important for setting alerts to family members or carers if an individual hasn't triggered a sensor in the morning, or is continuing to trigger sensors past their typical sleep time.  In contrast, warnings within a day are likely to be less useful as the routine across days can vary considerably more.

\subsection{Dynamic Functional Connectivity in Resting-state fMRI}
Functional connectivity (FC) is the area of neuroimaging research that studies patterns of interregional interactions in the brain. Resting-state functional magnetic imaging (rs-fMRI) data, which consists of time series data across spatial units (e.g., voxels, parcels, regions) can be used to noninvasively quantify FC. Static FC assumes stationarity in the data, and the simplest approach is to calculate the correlation matrix from the multivariate time series. However, there has been a rapid growth in interest in \textit{dynamic} FC (dFC), which assumes that FC changes over time \citep{lurie20}. There are several complementary analytic pipelines for estimating dFC, with each approach making different assumptions about the data. Sliding window analyses \citep{allen14} and hidden Markov models \citep{vidaurre17,fiecas23} are popular approaches for dFC analysis. \citet{lurie20} gave a more comprehensive overview of the approaches and challenges with dFC analyses.

We analysed data from the Brain Imaging Development of Girls' Emotion and Self (BRIDGES) Study, a longitudinal study whose aim is to understand brain development in adolescents with non-suicidal self-injury (NSSI). The BRIDGES Study recruited individuals 12-16 years of age who identified as female at birth and who exhibited a continuum of NSSI severity representing four categories: No NSSI, Mild NSSI, Moderate NSSI, and Severe NSSI \citep{nair23}. Our analyses focused on a cross-sectional assessment of the rs-fMRI data from $N=137$ adolescents. Detailed descriptions of data acquisition and preprocessing has been reported elsewhere \citep{basgoze21,basgoze23}.

Our goal was to carry out a dFC analysis, and determine how dFC differs across NSSI severity. Following our recent work demonstrating the relevance of the default network to suicidal risk in children \citep{wiglesworth23}, here we focused our analyses on brain regions from the core hubs of the default mode network, consisting of the anterior cingulate cortex (ACC), medial prefrontal cortex (mPFC), and posterior cingulate cortex (PCC). Specifically, Glasser Clusters 18 and 19 contain the PCC and ACC/mPFC, respectively \citep{glasser16}. We used these two Clusters from the left and right hemispheres of the brain to form $K=4$ groups, with each group containing 14 parcels, leading to time series data from $\sum_{i=1}^4 k_i = 56$ spatial locations. We demeaned each time series, and removed the autocorrelation by fitting an autoregressive model of order 8 to each time series and extracted the residuals. We then scaled the residual time series to have unit variance, and we use the resulting time series in our analyses. In summary, our time series data had length $n=904$, and $p=56$ dimensions that formed $K=4$ groups with $k_i=14$ for each $i = 1,\ldots, 4$, obtained from $N=137$ individuals. These individuals had NSSI classifications as follows: 40 No NSSI,  13  Mild NSSI, 51  Moderate NSSI, and 33 Severe NSSI.

\begin{figure}
    \centering
    \includegraphics[width=1\linewidth]{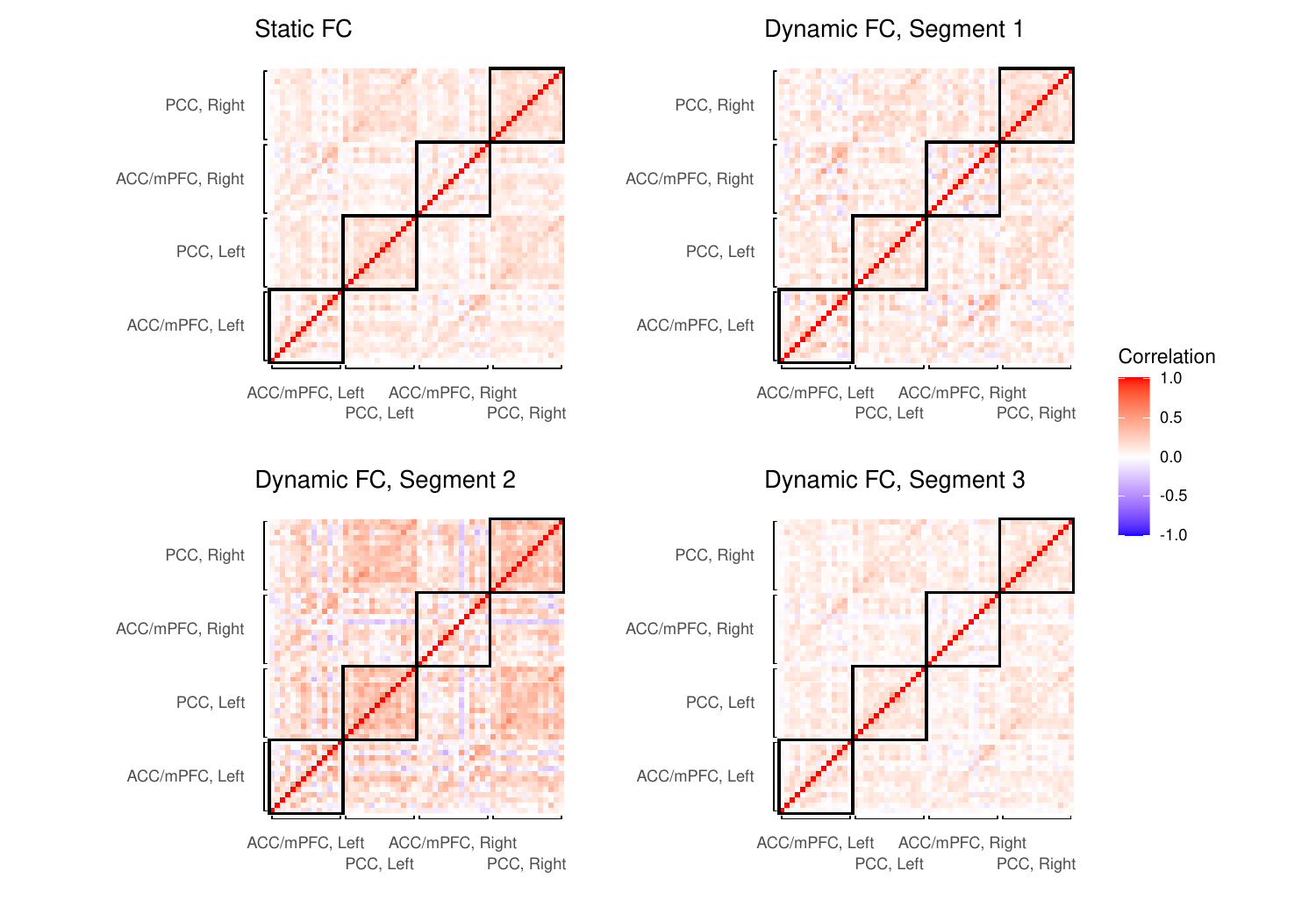}
    \caption{Results of the changepoint analysis using PELT-LMEC-HB from one individual. Two changepoints were detected, leading to three segments. The four matrices are correlation matrices, assuming no changepoints (Static FC) and also for each of the three segments (Dynamic FC). The black squares correspond to the regions of the brain: the anterior cingulate cortex (ACC) and medial prefrontal cortex (mPFC), and posterior cingulate cortex (PCC), for each of the left and right hemispheres. These four regions form the four groups in the analysis.}
    \label{fig:dfc}
\end{figure}

For each individual, we fit the PELT-LMEC-UB and PELT-LMEC-HB methods to their data and then we extracted the number of changepoints obtained from each method.  PELT-LMEC-UB detected 0 changepoints in 108 individuals. This could be due to the method being prone to false negatives, as shown in our simulations. PELT-LMEC-HB, on the other hand, detected a range of 0 to 5 changepoints across all individuals. Figure \ref{fig:dfc} shows the results from one individual for which PELT-LMEC-HB detected 2 changepoints, giving rise to three segments and hence three correlation matrices. These three correlation matrices across the three segments captures the temporal dynamics of functional connectivity between and within these brain regions for this specific individual. We can also see in Figure \ref{fig:dfc} the ``static'' functional connectivity for this individual, which is the correlation matrix of the data for this person assuming no changepoints. In contrast to static FC, dFC captures the changes in the correlations over time, for instance,  stronger correlations were present during Segment 2 and weaker correlations were present during Segment 3. 

In addition, Figure \ref{fig:dfc} gives us a qualitative look at the block structure in the correlation matrices, both when assuming no changepoints and also within each segment after estimating the changepoints. The subregions within the PCC, in particular, are strongly inter-correlated, and the PCC also shows strong between-hemisphere correlations. In contrast, the ACC/mPFC has greater variability of the strength in the correlations between its subregions, and its between-hemisphere block structure is not as strong as the PCC. Not shown in the figure are the variances of each dimension of the time series. For Segment 1, the variances for the ACC/mPFC Left, PCC Left, ACC/mPFC Right, and PCC Right, had ranges (0.84, 1.22), (0.84, 1.08), (0.81, 1.22), and (0.92, 1.15), respectively, thereby providing justification for the heterogeneous block assumption.

To determine if this measure of dynamic functional connectivity has clinical relevance in this sample, we carried out a posthoc analysis on the number of changepoints using Poisson regression to determine the association between NSSI severity and the number of changepoints. We set the the number of changepoints as the outcome NSSI severity group as the predictor, with No NSSI as the reference group. Results from the Poisson regression indicated a higher mean for the number of changepoints for those with Mild, Moderate, and Severe NSSI relative to those with No NSSI, and the increase in the number of changepoints for both Moderate and Severe NSSI relative to No NSSI was statistically significant (p-values = 0.017 and 0.023, respectively). Given recent literature on the impact of motion on dynamic connectivity analyses \citep{laumann17}, we also calculated each individual’s framewise displacement (FD), which is a measure of motion at each time point; greater values of FD correspond to greater motion. We counted the number of time points that each individual’s FD exceeded 0.3mm, and included this number as a regressor in the Poisson regression. After adjusting for FD and age, the direction of the effects were the same, but only the increase in the number of changepoints for those with Severe NSSI relative to No NSSI was statistically significant (p-value = 0.033). Altogether, these results suggest greater temporal dynamics in the functional connectivity within the default mode network as NSSI severity worsens.

%\begin{figure}
%\centering
%\subfigure[An example of an individual figure sub-caption.]{%
%\resizebox*{5cm}{!}{\includegraphics{graph1.eps}}}\hspace{5pt}
%\subfigure[A slightly shorter sub-caption.]{%
%\resizebox*{5cm}{!}{\includegraphics{graph2.eps}}}
%\caption{Example of a two-part figure with individual sub-captions
% showing that captions are flush left and justified if greater
% than one line of text.} \label{sample-figure}
%\end{figure}

\section{Discussion}\label{sec:discussion}
% Summary
We used the mixed model framework to develop novel methods for changepoint analysis for time series data. The mixed model framework is flexible, which allowed us to model univariate time series from a potentially non-normal distribution, and model the covariance matrix of multivariate time series. We embedded the mixed model in PELT for changepoint detection, giving rise to our methods that are flexible with respect to data type and are computationally fast. Through extensive simulations, we showed the gains in performance of our methods relative to existing methods, and showed how our methods perform in challenging scenarios. Finally, we illustrated the flexibility of our methods through two distinct applications.

% Possible extensions
The flexibility of the mixed model framework allows for natural extensions of our methods. For PELT-LMEC, we focused on covariance matrices with the UB or HB structure. \citet{yang24} further developed methods for estimating a precision matrix whenever the covariance matrix had a UB structure, which may be adapted for the HB structure and into PELT-LMEC. This would allow for changepoints in the graphical structure of the data, as previously investigated by \citet{cribben13}. We only used the UB and HB structures, but one can use other structures or make other assumptions (e.g., sparsity) for the covariance (or precision) matrix, though one should be mindful of the number of parameters in the covariance matrix and the computational speed in estimating that structured covariance matrix.

% Limitations
Just as how our methods inherit the strengths of both the mixed modeling framework and PELT, they also inherit their limitations. For instance, there are parameters that must be set in advance, as described in Section \ref{sec:practical}, which will affect the method's performance. For instance, the combination of the minimum segment length and the penalty will limit the method's ability to detect changepoints whenever the dimensionality of the data exceeds the length of the time series. A potential solution is to make strong assumption about the data (e.g., the time series can be arranged to a small number of groups), or impose sparsity constraints in the mixed model parameters as described above.

% Closing remarks

%\begin{itemize}
%    \item Summary of contributions, relation to existing methods
    %\item Possible extensions
    %\item Limitations (impact of autocorrelation, dimensionality, possible misspecification in group structure...)
    %\item Closing remarks
%\end{itemize}

% We have developed a framework for identifying changepoints in different forms of random effect models for specific applications in digital healthcare.  This framework combines a mixed effects model fitting procedure with the PELT search algorithm to detect changes in the underlying mixed effects model over time.

\section*{Acknowledgement(s)}

% An unnumbered section, e.g.\ \verb"\section*{Acknowledgements}", may be used for thanks, etc.\ if required and included \emph{in the non-anonymous version} before any Notes or References.

% \section*{Funding}
% Fiecas acknowledges funding from $\ldots$.
Killick acknowledges funding from the UK Engineering and Physical Sciences Research Council (EPSRC) under grant numbers EP/T021020/1 and EP/T014105/1. This research was supported by grants awarded by the National Insitute of Mental Health (R01-MH122473 and R01-MH107394), with the support from the Center for Magnetic Resonance Research (NIBIB P41 EB027061) and the High Performance Connectome Upgrade for Human 3T MR Scanner (1S10OD017974-01).

\bibliographystyle{apalike}
\bibliography{bib}

\begin{thebibliography}{}

\bibitem[Allen et~al., 2014]{allen14}
Allen, E.~A., Damaraju, E., Plis, S.~M., Erhardt, E.~B., Eichele, T., and Calhoun, V.~D. (2014).
\newblock Tracking whole-brain connectivity dynamics in the resting state.
\newblock {\em Cerebral cortex}, 24(3):663--676.

\bibitem[Ba{\c{s}}g{\"o}ze et~al., 2023]{basgoze23}
Ba{\c{s}}g{\"o}ze, Z., Demers, L., Thai, M., Falke, C.~A., Mueller, B.~A., Fiecas, M.~B., Roediger, D.~J., Thomas, K.~M., Klimes-Dougan, B., and Cullen, K.~R. (2023).
\newblock A multi-level examination of cognitive control in adolescents with non-suicidal self-injury.
\newblock {\em Biological Psychiatry Global Open Science}.

\bibitem[Ba{\c{s}}g{\"o}ze et~al., 2021]{basgoze21}
Ba{\c{s}}g{\"o}ze, Z., Mirza, S.~A., Silamongkol, T., Hill, D., Falke, C., Thai, M., Schreiner, M.~W., Parenteau, A.~M., Roediger, D.~J., Hendrickson, T.~J., et~al. (2021).
\newblock Multimodal assessment of sustained threat in adolescents with nonsuicidal self-injury.
\newblock {\em Development and psychopathology}, 33(5):1774--1792.

\bibitem[Cho and Fryzlewicz, 2015]{Cho15}
Cho, H. and Fryzlewicz, P. (2015).
\newblock Multiple-change-point detection for high dimensional time series via sparsified binary segmentation.
\newblock {\em Journal of the Royal Statistical Society Series B: Statistical Methodology}, 77(2):475--507.

\bibitem[Cribben et~al., 2013]{cribben13}
Cribben, I., Wager, T.~D., and Lindquist, M.~A. (2013).
\newblock Detecting functional connectivity change points for single-subject fmri data.
\newblock {\em Frontiers in computational neuroscience}, 7:143.

\bibitem[Dominicus et~al., 2008]{Dominicus08}
Dominicus, A., Ripatti, S., Pedersen, N.~L., and Palmgren, J. (2008).
\newblock A random change point model for assessing variability in repeated measures of cognitive function.
\newblock {\em Statistics in medicine}, 27(27):5786--5798.

\bibitem[Fiecas et~al., 2023]{fiecas23}
Fiecas, M., Coffman, C., Xu, M., Hendrickson, T., Mueller, B., Klimes-Dougan, B., and Cullen, K. (2023).
\newblock Approximate hidden semi-markov models for dynamic connectivity analysis in resting-state fmri.
\newblock {\em Statistics and its Interface}, 16(2):259--277.

\bibitem[Franke et~al., 2012]{Franke12}
Franke, J., Kirch, C., and Kamgaing, J.~T. (2012).
\newblock Changepoints in times series of counts.
\newblock {\em Journal of Time Series Analysis}, 33(5):757--770.

\bibitem[Fryzlewicz, 2014]{Fryzlewicz14}
Fryzlewicz, P. (2014).
\newblock Wild binary segmentation for multiple change-point detection.
\newblock {\em The Annals of Statistics}, 42(6):2243--2281.

\bibitem[Gallagher et~al., 2022]{gallagher22}
Gallagher, C., Killick, R., Lund, R., and Shi, X. (2022).
\newblock Autocovariance estimation in the presence of changepoints.
\newblock {\em Journal of the Korean Statistical Society}, 51(4):1021--1040.

\bibitem[Gillam et~al., 2022]{HowzJessSIM}
Gillam, J., Killick, R., Heal, J., and Norwood, B. (2022).
\newblock Modeling and forecasting of at home activity in older adults using passive sensor technology.
\newblock {\em Statistics in Medicine}, 41(23):4629--4646.

\bibitem[Glasser et~al., 2016]{glasser16}
Glasser, M.~F., Coalson, T.~S., Robinson, E.~C., Hacker, C.~D., Harwell, J., Yacoub, E., Ugurbil, K., Andersson, J., Beckmann, C.~F., Jenkinson, M., et~al. (2016).
\newblock A multi-modal parcellation of human cerebral cortex.
\newblock {\em Nature}, 536(7615):171--178.

\bibitem[Killick et~al., 2012]{PELT}
Killick, R., Fearnhead, P., and Eckley, I.~A. (2012).
\newblock Optimal detection of changepoints with a linear computational cost.
\newblock {\em Journal of the American Statistical Association}, 107(500):1590--1598.

\bibitem[Lai and Albert, 2014]{Lai14}
Lai, Y. and Albert, P.~S. (2014).
\newblock Identifying multiple change points in a linear mixed effects model.
\newblock {\em Statistics in medicine}, 33(6):1015--1028.

\bibitem[Laumann et~al., 2017]{laumann17}
Laumann, T.~O., Snyder, A.~Z., Mitra, A., Gordon, E.~M., Gratton, C., Adeyemo, B., Gilmore, A.~W., Nelson, S.~M., Berg, J.~J., Greene, D.~J., et~al. (2017).
\newblock On the stability of bold fmri correlations.
\newblock {\em Cerebral cortex}, 27(10):4719--4732.

\bibitem[Li et~al., 2023]{Li23}
Li, Y.-N., Li, D., and Fryzlewicz, P. (2023).
\newblock Detection of multiple structural breaks in large covariance matrices.
\newblock {\em Journal of Business \& Economic Statistics}, 41(3):846--861.

\bibitem[Lund et~al., 2023]{Lund23}
Lund, R.~B., Beaulieu, C., Killick, R., Lu, Q., and Shi, X. (2023).
\newblock Good practices and common pitfalls in climate time series changepoint techniques: A review.
\newblock {\em Journal of Climate}, 36(23):8041--8057.

\bibitem[Lurie et~al., 2020]{lurie20}
Lurie, D.~J., Kessler, D., Bassett, D.~S., Betzel, R.~F., Breakspear, M., Kheilholz, S., Kucyi, A., Li{\'e}geois, R., Lindquist, M.~A., McIntosh, A.~R., et~al. (2020).
\newblock Questions and controversies in the study of time-varying functional connectivity in resting fmri.
\newblock {\em Network neuroscience}, 4(1):30--69.

\bibitem[Nair et~al., 2023]{nair23}
Nair, A.~U., Brekke-Riedl, J.~A., DiMaggio-Potter, M.~E., Carosella, K.~A., Lasch, C., Brower, R., Papke, V., Reigstad, K., Klimes-Dougan, B., and Cullen, K.~R. (2023).
\newblock Clinical trajectories in adolescents with and without a history of non-suicidal self-injury: the bridges longitudinal study.
\newblock {\em Journal of psychiatry and brain science}, 8.

\bibitem[Naumova et~al., 2001]{Naumova01}
Naumova, E.~N., Must, A., and Laird, N.~M. (2001).
\newblock Tutorial in biostatistics: evaluating the impact of ‘critical periods’ in longitudinal studies of growth using piecewise mixed effects models.
\newblock {\em International journal of epidemiology}, 30(6):1332--1341.

\bibitem[Pignatiello~Jr and Samuel, 2001]{Pignatiello01}
Pignatiello~Jr, J.~J. and Samuel, T.~R. (2001).
\newblock Identifying the time of a step-change in the process fraction nonconforming.
\newblock {\em Quality Engineering}, 13(3):357--365.

\bibitem[Ryan and Killick, 2023]{ryan23}
Ryan, S. and Killick, R. (2023).
\newblock Detecting changes in covariance via random matrix theory.
\newblock {\em Technometrics}, 65(4):480--491.

\bibitem[Shi et~al., 2022]{Shi22}
Shi, X., Gallagher, C., Lund, R., and Killick, R. (2022).
\newblock A comparison of single and multiple changepoint techniques for time series data.
\newblock {\em Computational Statistics \& Data Analysis}, 170:107433.

\bibitem[Taylor et~al., 2021a]{Taylor21}
Taylor, S.~A., Killick, R., Burr, J., and Rogerson, L. (2021a).
\newblock Assessing daily patterns using home activity sensors and within period changepoint detection.
\newblock {\em Journal of the Royal Statistical Society Series C: Applied Statistics}, 70(3):579--595.

\bibitem[Taylor et~al., 2021b]{HowzSimonRSSC}
Taylor, S. A.~C., Killick, R., Burr, J., and Rogerson, L. (2021b).
\newblock {Assessing Daily Patterns Using Home Activity Sensors and Within Period Changepoint Detection}.
\newblock {\em Journal of the Royal Statistical Society Series C: Applied Statistics}, 70(3):579--595.

\bibitem[Truong et~al., 2020]{Truong20}
Truong, C., Oudre, L., and Vayatis, N. (2020).
\newblock Selective review of offline change point detection methods.
\newblock {\em Signal Processing}, 167:107299.

\bibitem[Vidaurre et~al., 2017]{vidaurre17}
Vidaurre, D., Smith, S.~M., and Woolrich, M.~W. (2017).
\newblock Brain network dynamics are hierarchically organized in time.
\newblock {\em Proceedings of the National Academy of Sciences}, 114(48):12827--12832.

\bibitem[Wiglesworth et~al., 2023]{wiglesworth23}
Wiglesworth, A., Falke, C.~A., Fiecas, M., Luciana, M., Cullen, K.~R., and Klimes-Dougan, B. (2023).
\newblock Brain signatures in children who contemplate suicide: learning from the large-scale abcd study.
\newblock {\em Psychological medicine}, 53(5):2164--2173.

\bibitem[Wu et~al., 2021]{Wu21}
Wu, Q., Ma, T., Liu, Q., Milton, D.~K., Zhang, Y., and Chen, S. (2021).
\newblock Icn: extracting interconnected communities in gene co-expression networks.
\newblock {\em Bioinformatics}, 37(14):1997--2003.

\bibitem[Yang et~al., 2024]{yang24}
Yang, Y., Chen, C., and Chen, S. (2024).
\newblock Covariance matrix estimation for high-throughput biomedical data with interconnected communities.
\newblock {\em The American Statistician}, (just-accepted):1--19.

\bibitem[Zeileis et~al., 2002]{zeileis02}
Zeileis, A., Leisch, F., Hornik, K., and Kleiber, C. (2002).
\newblock strucchange: An r package for testing for structural change in linear regression models.
\newblock {\em Journal of statistical software}, 7:1--38.

\bibitem[Zheng et~al., 2022]{PELTconsistency}
Zheng, C., Eckley, I., and Fearnhead, P. (2022).
\newblock {Consistency of a range of penalised cost approaches for detecting multiple changepoints}.
\newblock {\em Electronic Journal of Statistics}, 16(2):4497 -- 4546.

\end{thebibliography}

%\appendix
%\section{Appendices}

\end{document}